\documentclass[journal,transmag]{IEEEtran}
\usepackage[utf8]{inputenc}
\usepackage{glossaries-prefix}
\usepackage{amsmath}
\usepackage{amssymb}
\usepackage{todonotes}
\usepackage[square,numbers,comma]{natbib}
\usepackage{gensymb}
\usepackage{hyperref}
\usepackage[caption=false]{subfig}
\usepackage{multirow}
\usepackage{booktabs}
\usepackage{siunitx}
\usepackage{tikz}
\newcommand{\ra}[1]{\renewcommand{\arraystretch}{#1}}

\newacronym{pid}{PID}{proportional-integral-derivative}
\newacronym[prefixfirst={a\ },% prefix used on first use
            prefix={an\ }% prefix used on subsequent use
            ]{ml}{ML}{machine learning}
\newacronym{drl}{DRL}{deep reinforcement learning}
\newacronym[prefixfirst={a\ },% prefix used on first use
            prefix={an\ }% prefix used on subsequent use
            ]{nn}{NN}{neural network}
\newacronym[prefixfirst={a\ },% prefix used on first use
            prefix={an\ }% prefix used on subsequent use
            ]{rnn}{RNN}{recurrent neural network}
\newacronym{cnn}{CNN}{convolutional neural network}
\newacronym[prefixfirst={a\ },% prefix used on first use
            prefix={an\ }% prefix used on subsequent use
            ]{rl}{RL}{reinforcement learning}
\newacronym[prefixfirst={a\ },% prefix used on first use
            prefix={an\ }% prefix used on subsequent use
            ]{mdp}{MDP}{Markov decision process}
\newacronym[prefixfirst={a\ },% prefix used on first use
            prefix={an\ }% prefix used on subsequent use
            ]{mpc}{MPC}{Model Predictive Control}
\newacronym{ddpg}{DDPG}{deep deterministic policy gradients}
\newacronym{td3}{TD3}{twin delayed DDPG}
\newacronym{sac}{SAC}{Soft Actor Critic}
\newacronym{her}{HER}{hindsight experience replay}
\newacronym{brl}{BRL}{batch reinforcement learning}
\newacronym{uav}{UAV}{unmanned aerial vehicle}
\newacronym{ppo}{PPO}{proximal policy optimization}
\newacronym{caps}{CAPS}{conditioning for action policy smoothness}
\newacronym{fbwa}{fbwa}{fly-by-wire-A}
\newacronym[prefixfirst={a\ },% prefix used on first use
            prefix={an\ }% prefix used on subsequent use
            ]{fc}{FC}{fully-connected}
\newacronym{sbc}{SBC}{Single-board computer}
\newacronym{adp}{ADP}{adaptive dynamic programming}
\newacronym{lqr}{LQR}{linear quadratic regulator}
\newacronym{hjb}{HJB}{Hamilton–Jacobi–Bellman}

\newcommand\copyrighttext{%
  \footnotesize \textcopyright 2023 IEEE. Personal use of this material is permitted.
  Permission from IEEE must be obtained for all other uses, in any current or future
  media, including reprinting/republishing this material for advertising or promotional
  purposes, creating new collective works, for resale or redistribution to servers or
  lists, or reuse of any copyrighted component of this work in other works.
  DOI: \href{https://doi.org/10.1109/TNNLS.2023.3263430}{10.1109/TNNLS.2023.3263430}}
\newcommand\copyrightnotice{%
\begin{tikzpicture}[remember picture,overlay]
\node[anchor=south,yshift=6pt] at (current page.south) {\fbox{\parbox{\dimexpr\textwidth-\fboxsep-\fboxrule\relax}{\copyrighttext}}};
\end{tikzpicture}%
}

\DeclareMathOperator*{\argmax}{argmax} % no space, limits underneath in displays

\title{Data-Efficient Deep Reinforcement Learning for Attitude Control of Fixed-Wing UAVs: Field Experiments}

\author{\IEEEauthorblockN{Eivind Bøhn,
Erlend M. Coates,
Dirk Reinhardt, and
Tor Arne Johansen}% <-this % stops an unwanted space
\thanks{%Manuscript received December 1, 2012; revised August 26, 2015. 
Eivind Bøhn is with the Department of Mathematics and Cybernetics, SINTEF DIGITAL, Oslo, Norway (e-mail: eivind.bohn@sintef.no)

Erlend M. Coates, Dirk Reinhardt, and Tor Arne Johansen are with the Centre for Autonomous Marine Operations and Systems, Department of Engineering Cybernetics, NTNU, Trondheim, Norway (e-mail: \{erlend.coates, dirk.reinhardt, tor.arne.johansen\}@ntnu.no)
This research was funded by the Research Council of Norway through PhD Scholarships at SINTEF, grant number 272402, and  through the Centres of Excellence funding scheme, grant number 223254 NTNU AMOS, and grant number 261791 Autofly.}}

\begin{document}

\maketitle
\copyrightnotice
%\markboth{Please submit the manuscript to the Special Issue on Reinforcement Learning Based Control: Data-Efficient and Resilient Methods}{}

\begin{abstract}
Attitude control of fixed-wing \glspl{uav} is a difficult control problem in part due to uncertain nonlinear dynamics, actuator constraints, and coupled longitudinal and lateral motions. Current state-of-the-art autopilots are based on linear control and are thus limited in their effectiveness and performance. \Gls{drl} is a machine learning method to automatically discover optimal control laws through interaction with the controlled system, which can handle complex nonlinear dynamics. We show in this paper that \gls{drl} can successfully learn to perform attitude control of a fixed-wing \gls{uav} operating directly on the original nonlinear dynamics, requiring as little as three minutes of flight data. We initially train our model in a simulation environment and then deploy the learned controller on the \gls{uav} in flight tests, demonstrating comparable performance to the state-of-the-art ArduPlane \gls{pid} attitude controller with no further online learning required. Learning with significant actuation delay and diversified simulated dynamics were found to be crucial for successful transfer to control of the real \gls{uav}. In addition to a qualitative comparison with the ArduPlane autopilot, we present a quantitative assessment based on linear analysis to better understand the learning controller's behavior.
\end{abstract}\vspace{-0.5cm}


\glsresetall

\section{Introduction}\label{sec:introduction}
\begin{figure}[ht]
    \centering
    \includegraphics[width=0.485\textwidth]{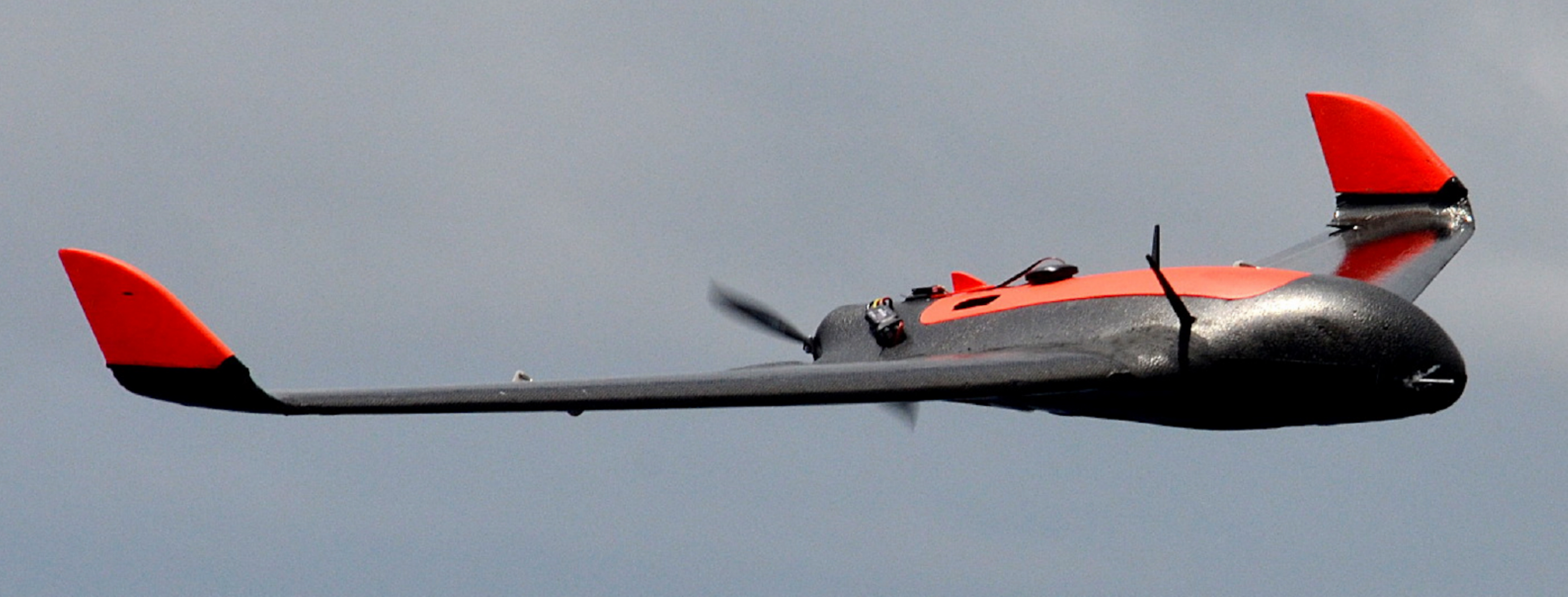}
    \caption{The Skywalker X8 fixed-wing aircraft in flight.}
    \label{fig:x8}
\end{figure}

Many challenging control problems arise during advanced operation of fixed-wing \glspl{uav}, such as aerobatic maneuvering~\cite{bulka2019}, perching~\cite{cory2008}, deep-stall landing~\cite{mathisen2021}, recovery from loss of control~\cite{cunis2020}, flying in strong wind fields~\cite{liu2016}, or performing VTOL transitions between hover and forward flight~\cite{anglade2019}. Fixed-wing \glspl{uav}, as illustrated in Fig.~\ref{fig:x8}, have superior range and endurance when compared to multirotor \glspl{uav}. However, the control of such vehicles is challenging due to highly coupled, underactuated nonlinear dynamics, as well as uncertain aerodynamics affected by wind disturbances that make up a large fraction of the vehicle's airspeed.

A class of methods that have shown promising results for challenging control problems is \gls{drl}. \Gls{rl} is an area of machine learning concerned with learning optimal sequential decision-making. \Gls{drl} is the combination of \gls{rl} algorithms with \glspl{nn} as function approximators, which is the state-of-the-art approach for many problems requiring complex decision making over long time horizons such as game-playing \cite{schrittwieser2020mastering}, dexterous in-hand robotic manipulation \cite{andrychowicz2020learning}, and object manipulation \cite{schoettler2020deep}. It can handle continuous state and action spaces, highly complex and nonlinear system dynamics, and in general, does not require a model of the system to be controlled. Furthermore, the online execution of an \gls{rl} controller is often very computationally efficient. This should make \gls{drl} an enticing alternative for problems where accurate identification of the system is difficult and traditional control approaches are unable to yield sufficient control performance. Despite this potential, \gls{drl} has yet to be widely adopted for control and notably lacks demonstrations of control applications outside of simulations. One of the main contributing factors to this is the lack of safety guarantees and the ability to formulate operating constraints, both in the exploration and exploitation phase, which is further complicated by the data-intensive nature of \gls{drl}. We will in the rest of this paper use \gls{rl} to refer to \gls{drl}.

An approach to mitigate the challenges of \gls{rl} for control is foregoing online exploration entirely and learning the controller exclusively from historical data with no further interaction with the system to be optimized, i.e. offline \gls{rl} \cite{levine2020offline}. The latter is a radical alteration of the \gls{rl} problem introducing new challenges and necessitating its own set of learning algorithms. It could nevertheless be an important tool in the future for problems such as control of \glspl{uav} --- where data collection carries a high risk and accurate modeling is difficult. A more common approach is performing part of or the whole exploration phase in a simulation of the target system. A downside of this approach is that while \gls{rl} in principle is a model-free optimization framework, the success of the transfer from the simulated environment to the real target environment is highly affected by the accuracy of the simulation model, the lack of which is referred to as the reality gap in \gls{rl}. One should therefore take great effort in minimizing the reality gap through sim-to-real measures, which aim at robustifying the learned controller and emulating effects such as latency and measurement noise present in the real control system. For a recent survey on sim-to-real methods in the context of control and robotics, see \cite{zhao2020sim}.

For the sim-to-real learning approach to be useful for practical flight control applications, controllers trained in simulation need to transfer well to control the real \gls{uav}. Before attempting advanced problems like e.g. deep-stall landing, it makes sense to first attempt simpler problems and identify what factors are important for controllers to transfer well to reality. In this paper, we consider the attitude control problem of fixed-wing \glspl{uav}. Attitude refers to the orientation of the aircraft, and control of the attitude constitutes the lowest level of flight control deciding how the actuators of the \gls{uav} are used to achieve the desired attitude as decided by the guidance components of the control system. This work is a follow-up to our previous work \cite{bohn2019deep} in which we demonstrate the efficacy of \gls{drl} for attitude control of fixed-wing \glspl{uav} in a simulator environment. We now target control of the real \gls{uav} in the field and develop a framework to learn attitude controllers with a focus on data efficiency, yielding flightworthy controllers with only minutes of learning time. Starting with a \gls{uav} model obtained primarily through wind-tunnel experiments, we adopt the method of exploring and learning in a simulator environment and iteratively adjust the model and simulation environment with insights from flight experiments. We extensively apply domain randomization and other sim-to-real measures in order to reduce the reality gap. Moreover, the data efficiency of our method limits overfitting to the simulation model, such that the controller transfers better to the field, and when combined with safe exploration measures the high data efficiency could enable learning controllers entirely on the real \gls{uav} in the field.

The literature on \gls{rl}-based attitude control of \glspl{uav} is dominated by quadcopters, and most works operate exclusively in simulated environments \cite{koch2019reinforcement,lin2019supplementary,wang2020deterministic,fei2020learn,bernini2021few}. When it comes to works presenting real-world flight experiments we have identified the following: \cite{lambert2019low,xu2019learning,koch2019neuroflight,clarke2020closed,fletcher2021reinforcement,wada2021unmanned,mysore2021train}. Of these, only \cite{clarke2020closed}, their follow-up work \cite{fletcher2021reinforcement}, and \cite{wada2021unmanned} use a fixed-wing \gls{uav} design. \cite{clarke2020closed} and \cite{fletcher2021reinforcement} study the specific problem of controlling a perched landing, \cite{wada2021unmanned} fixes the aircraft in a wind-tunnel and limit themselves to controlling the pitch of the \gls{uav}, while we study the attitude control problem of an unconstrained vehicle in a 3D outdoor environment. Moreover, the data requirement of their methods is on the order of millions of samples. The other aforementioned works also require millions of samples of data, with the notable exception of \cite{lambert2019low}. Their method uses model-based \gls{rl} and involves learning a forward dynamics model that is used in \pgls{mpc} scheme which controls the quadrotor. While this method is very sample efficient, the \gls{mpc} is too computationally complex to run aboard the \gls{uav} and therefore necessitates continuous communication with an external computer, while our controller can run on a fraction of the computational power embedded in the \gls{uav}. For a more general overview of the application of \gls{rl} to \glspl{uav} see \cite{azar2021drone} and the related works section of \cite{bohn2019deep}.

Other related control methods that have been proposed for attitude control of fixed-wing \glspl{uav} include simple \gls{pid} loops~\cite{Beard2012}, the \gls{lqr}~\cite{lavretsky2013}, \gls{adp}~\cite{Ferrari2004,Zhou2017,Zhang2022}, and \gls{mpc}~\cite{Reinhardt2022thesis}. \gls{adp} is similarly to \gls{rl} a data-driven optimal control scheme. In the works of \cite{Ferrari2004,Zhou2017,Zhang2022}, the cost function is assumed to have a quadratic form and the optimal control law is derived from the \gls{hjb} equation, which is solved numerically by approximating the cost-function using a value iteration scheme. Of note is also \gls{mpc} which has inherent support for system constraints, multivariate objectives, high interpretability, and the ability to incorporate domain knowledge in the dynamics model. Motivated by these properties, the authors of this work have developed an \gls{mpc} controller for simultaneous control of the \gls{uav}'s attitude and airspeed, allowing the controller to consider the coupling between pitch and airspeed dynamics (note that we have previously demonstrated that \gls{rl} is also capable of learning this coupled control problem \cite{bohn2019deep,bohn2022thesis}). Flight experiments with this \gls{mpc} were conducted using the same experimental platform as this work, albeit with a more powerful processing unit. See \cite{Reinhardt2022thesis} for details and experimental results, as well as the discussion in~\cite{coates2022}. Due to the differences in control objective and hardware platform, an in-depth comparison between \gls{mpc} and the \gls{rl} controller is outside the scope of this paper.

The contributions and novelty of this paper can be summarized as follows:

\begin{itemize}
    \item To the best of our knowledge, this is the first work to demonstrate through field experiments the efficacy of \gls{rl} for attitude control of fixed-wing \glspl{uav}, a class of \gls{uav} design generally considered to be significantly more complex to control than the multirotor which is common in the literature.
    \item The proposed method improves upon the data efficiency of the existing literature by at least an order of magnitude. We show that our method can develop flightworthy controllers with only 3 minutes of data from interaction with the simulation environment, providing an important step towards enabling the learning of \gls{rl} controllers entirely on the real \gls{uav}. 
    \item We present an analysis of the \gls{rl} controller in order to better understand how it operates, including a comparison to an industry-standard \gls{pid} controller. 
\end{itemize}

The rest of this paper is organized as follows: Section~\ref{sec:preliminaries} presents the \gls{rl} framework used to optimize the controller. Section~\ref{sec:method} describes our method in detail and presents our thoughts on what parameters are important for the learning problem. Section~\ref{sec:experiments} presents the main experimental results obtained in real flight, followed by a discussion and analysis part in Section~\ref{sec:discussion}. Finally, Section~\ref{sec:conclusion} concludes with our thoughts on the presented work and ideas for future research. \markboth{}{}
\section{Reinforcement Learning} \label{sec:preliminaries}
The \gls{rl} optimization framework consists of two main parts, an environment that is to be controlled, and an agent that observes the state of the environment and selects actions to maximize the rewards it receives. The environment is typically described as \pgls{mdp}, which is defined by a 5-tuple of components $(\mathcal{S}, \mathcal{A}, \mathcal{T}, R, \gamma)$: A set of states, $\mathcal{S}$, a set of actions available to the agent, $\mathcal{A}$, a transition function $\mathcal{T}(s_t, a_t) = s_{t+1}$ which describe the evolution of the states as a function of actions and previous states, a reward function $R(s, a)$ quantifying the utility of states and accompanying actions, and finally, a discount factor $\gamma \in [0, 1)$, weighing the relative value of immediate and future rewards. %Since we in this work use \gls{rl} for the purpose of control, we consider an extension to the \gls{mdp} in the form of a goal state $g$ from a set of goal states $\mathcal{G}$, that defines the desired state the agent is to control the environment towards. 

We consider in this paper the episodic finite-horizon formulation of \gls{rl}. An episode is a trajectory of states and actions $\tau = (s_0, a_0, s_1, a_1, \dots, s_T)$ of maximum length $T$ with a distribution of initial states $\mathcal{S}_0$. The policy $\pi_\vartheta$ is a parameterized function that maps states to actions, describing the agent's behaviour (analogous to a control law in control terminology). The \gls{rl} objective can then be stated as finding the optimal parameters $\vartheta$ of the policy $\pi_\vartheta^*$ that maximizes the return $G$ over the distribution of the initial conditions of the episode:

\begin{align}
    G(\tau) &= \sum_{t=0}^T \gamma^t R(s_t, a_t) \label{eq:rl:return} \\
    \pi_\vartheta^* &= \argmax_{\vartheta} \mathbb{E}_{\tau \sim \mathcal{T}\left(\mathcal{S}_0, \pi_\vartheta\right)}\left[G(\tau)\right] \label{eq:rl:obj}
\end{align}

where $\sim$ signifies that the trajectories $\tau$ are sampled from $\mathcal{T}(\mathcal{S}_0, \pi)$, i.e. the distribution of trajectories given by the transition dynamics $\mathcal{T}$, the initial state distribution $\mathcal{S}_0$, and the action-distribution of the policy $\pi$.

\gls{sac} \cite{haarnoja2018soft} is an actor-critic algorithm whose defining characteristic is its entropy-regularization, meaning that it is jointly maximizing the expected rewards as in \eqref{eq:rl:obj} and maximizing the entropy of the policy:

\begin{align}
    &\pi_\vartheta^* = \argmax_{\vartheta} \mathbb{E}_{\tau \sim \mathcal{T}\left(\mathcal{S}_0, \pi_\vartheta\right)}\left[\sum_{t=0}^{T} \gamma^t \left( R(s_t, a_t) + \chi \mathcal{H}(\vartheta | s_t, a_t)\right)\right] \\
    &\vartheta_\mathrm{new} = \vartheta_{\mathrm{old}} + \eta \nabla_\vartheta J^{\mathrm{SAC}}(\pi_\vartheta) \label{eq:sac:ga}
\end{align}

where $\mathcal{H}(\vartheta | s, a) = \mathbb{E}_{a \sim \pi_\vartheta(a |s)}\left[-\log \pi_\vartheta(a | s)\right]$ is the entropy, equal to the negative log probability of the action-distribution of the policy in the state in question, $\chi$ is the entropy coefficient, and \eqref{eq:sac:ga} shows the gradient ascent scheme used to arrive at the optimal policy in which $\eta > 0$ is the learning rate and $J^{\mathrm{SAC}}$ is the \gls{sac} objective function. For brevity, we limit our discussion about \gls{sac} to the implementation of the policy and instead refer the reader to the original paper \cite{haarnoja2018soft} for details on the objective function. The policy is implemented as follows:

\begin{align}
    \pi_\vartheta(s_t) = \tanh(\mu_\vartheta(s) + \sigma_\vartheta(s) \odot \xi), \enspace \xi \sim \mathcal{N}(0, I) \label{eq:sac:pi}
\end{align}

here, $\mu_\vartheta$ and $\sigma_\vartheta$ are two parameterized deterministic functions of the input, representing the mean and covariance of the output, respectively. The notation $\odot$ denotes element-wise matrix-multiplication and $\xi$ is independently sampled Gaussian noise. The entropy can therefore be controlled in a state-dependent manner through the $\sigma_\vartheta$ function. Finally, the output is saturated with the hyperbolic tangent function, which squashes the Gaussian's infinite support to the domain $[-1, 1]$, limiting the adverse effects of extreme noise values and giving bounded outputs. During field experiments we set $\sigma_\vartheta = 0$ as this tends to yield better performance and smoother outputs \cite{haarnoja2018soft}.

\section{Method}\label{sec:method}
The control objective of the \gls{rl} controller is to control the attitude of the aircraft to the desired reference attitude. We use standard aircraft nomenclature and coordinate systems~\cite{Beard2012}, as well as a roll-pitch-yaw Euler angle parameterization of attitude. The heading/yaw angle is typically not controlled directly, but rather through banked-turn maneuvers~\cite{Beard2012}. Therefore, the natural choice of controlled states are the roll angle $\phi$, and the pitch angle $\theta$. We assume that the \gls{uav} is equipped with control surfaces such that the roll and pitch angles are controllable (an assumption that is satisfied by most \gls{uav} designs). The Skywalker X8 seen in Fig.~\ref{fig:x8} is used in our field experiments. It has two elevon control surfaces, one on each wing, which can be driven together to produce a pitching moment, or driven differentially to produce a rolling moment. In addition, it has a propeller that can produce a thrust force along the longitudinal axis of the \gls{uav}. In the simulation environment, a PI-controller is used to control airspeed using the propeller throttle~\cite{bohn2019deep}.

As a general approach, we tested new ideas in the simulation environment and made extensive use of sim-to-sim experiments where we studied how the controller transferred from simulation with one set of parameters to a simulation with another set of parameters. We then tested the most promising controllers in flight experiments in the field and adjusted our approach based on the insight we gathered from the flight experiments. The simulation environment software is made open-source and is available at \cite{repo}.

\subsection{Controller Architecture and State Design}
We identified in our previous work that limiting the state vector to only the most useful information and reducing redundancy is important for the rate of convergence, and to prevent the controller from learning spurious relationships. This has also been observed in other research \cite{bernini2021few}. At every time step we measure the following information:

\begin{align}
    \begin{split}
     m_t &= \big[p_t, q_t, r_t, \alpha_t, \beta_t, V_{a,t}, \delta_{r, t-1}, \delta_{l, {t-1}}, \\ 
     &\quad \enspace \> I_{\phi,t}, I_{\theta,t}, \phi_t, \theta_t, e_{\phi,t}, e_{\theta, t} \big]^\top \label{eq:measurement}   
    \end{split} \\
    I_{*,t} &= \gamma^I I_{*,t-1}  + e_{*,t}, \enspace \gamma^I = 0.99, \enspace I_0 = 0, * \in \{\phi, \theta \}
\end{align}

where $t$ is the time index, $\omega_t = \left[p_t, q_t, r_t\right]^\top$ is the angular velocity in the body-fixed frame, $\alpha_t$ is the angle-of-attack, $\beta_t$ is the sideslip-angle, $V_{a,t}$ is the airspeed, $\delta_{\{r, l\}, t-1}$ represent the previous output of the \gls{rl} controller, in this case the commanded deflection angles of the right and left elevons, $e_{*, t} = *_t - *_{r, t}$ is the state tracking error where subscript $r$ denotes the state reference, $I_*$ is the integrator of the state error and $\gamma^I$ is the integrator decay rate. The integration decay scheme follows \cite{xu2019learning}, and facilitates boundedness of the integrator state.
Lastly, because \glspl{nn} are known to converge faster given normalized inputs, the measurements are normalized using running estimates of mean and variance for each component before being fed to the controller.

\begin{figure}
    \centering
    \includegraphics[width=0.5\textwidth]{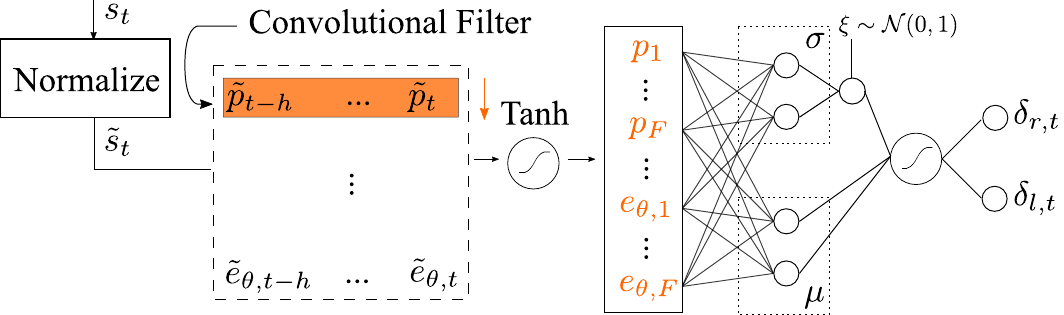}
    \caption{Architecture of the \gls{rl} controller, superscript $\sim$ signifies normalized states.}
    \label{fig:rlc_arc}
\end{figure}

Due to unmeasured effects such as turbulence and the sim-to-real measures described in Section~\ref{sec:method:s2r}, the attitude control problem is partially observable. Furthermore, to enable the controller to adapt to the dynamics of the field experiments, we wish to enhance the controller with the capability of inferring the dynamics around the current state. A common approach to achieve this effect is to use \pgls{rnn} \cite{wada2021unmanned,peng2018sim}. However, we found that using a one-dimensional convolution over the time dimension as the input layer yielded similar control performance, and therefore prefer it since it is significantly less complex than the \gls{rnn}. We therefore include the $h$ last measurements \eqref{eq:measurement} in the state vector, where $\hat{m}_t$ indicates a noisy measurement to be defined in Section~\ref{sec:method:s2r}:

\begin{align}
    s_t &= \left[\hat{m}_t, \hat{m}_{t-1}, \dots, \hat{m}_{t-h}\right]^\top \\
    \left[\delta_{r,t}, \delta_{l,t}\right]^\top &= \pi_\vartheta(s_t) + \left[\delta_{r, \mathrm{trim}}, \delta_{l, \mathrm{trim}}\right]^\top \label{eq:pi:output} 
\end{align}

such that the total size of the state vector is $|s_t| = |m_t| \cdot h$. The convolutional input layer scales favorably in number of learned parameters compared to \pgls{fc} layer: it scales linearly in $|m_t|$ as opposed to multiplicative for the \gls{fc} layer, and it is constant for $h$. The convolutional layers' output size is $F \cdot |m_t|$ where $F$ is the number of learned convolutional filters, and each filter has size $h$. The memory capacity of the state vector can therefore be increased as required to give sufficient history to infer the dynamics, with only a slight increase in the number of parameters. Through a small grid search using rate of learning and asymptotic performance as metrics we found $F=8$ and $h = 10$ to work well. The complete \gls{rl} controller architecture is shown in Fig. \ref{fig:rlc_arc}.

The output of the \gls{rl} controller is the commanded states of the controlled system's actuators relative to the trim-point \eqref{eq:pi:output}. The nominal elevon deflection angles $\delta_{r, \mathrm{trim}} = \delta_{l, \mathrm{trim}} = 0.045$ are calculated using a standard trim routine for horizontal, wings-level flight based on the model in Section~\ref{sec:method:uavmodel}~\cite{Beard2012}. The target \gls{uav} for the field experiments, the Skywalker X8, has elevon actuators and we therefore chose to have the controller output the desired deflection angles of these directly, in order to provide \gls{rl} with as direct control as possible. This choice is fairly arbitrary, however, and experiments showed that outputting virtual elevator and aileron angles (the sum and difference, respectively, of the elevon angles defined by~\eqref{eq:linmap1}-\eqref{eq:linmap2}) instead yield similar performance.

\subsection{Reward and Objective Design}\label{sec:method:reward}
We found sparse rewards to yield better results than shaped rewards, both in terms of rate of learning and in terms of asymptotic performance. A sparse reward is one that is nonzero only for some subset of the state space. It has the benefit that it is easier to formulate than hand-crafted shaped rewards, and would therefore be more transferable to other \glspl{uav} with fewer adjustments necessary. The reward is formulated as follows:

\begin{align}
    R(s_t, a_t) &= \lambda^\phi B(e_{\phi,t}) + \lambda^\theta B(e_{\theta,t}) + \lambda^{\dot{\phi}} B(\dot{\phi}_t) + \lambda^{\dot{\theta}} B(\dot{\theta}_t) \label{eq:method:reward} \\
    B(\cdot) &= \begin{cases}
                1 &\text{ if } \left|\cdot\right| \leq \cdot^b \\
                0 &\text{ otherwise}
            \end{cases} \\
    e_{\phi_{t}}^b &= 3 \degree, \enspace e_{\theta_{t}}^b = 3 \degree, \enspace \dot{\phi}_{t}^b = 4.3 \degree /s, \enspace \dot{\theta}_{t}^b = 4.3 \degree /s \\
    \lambda^{\phi} &= 0.5, \enspace \lambda^{\theta} = 0.5, \enspace \lambda^{\dot{\phi}} = 0.167, \enspace \lambda^{\dot{\theta}} = 0.167
\end{align}

where superscript $b$ refers to the goal-bound on the variable and the $\lambda$s are weighting factors. This reward ensures that the controller tracks the setpoints with accuracy as specified by the bound, while the rewards on the derivatives of the controlled states discourage high rates. Our method is not very sensitive to the size of the bound, but generally larger bounds accelerate learning at the expense of tracking accuracy.

When transferring from a simulator environment to the field, it is important to consider how the actuation system impacts the effects of actions. That is, while high-gain bang-bang control may be an optimal strategy in the simulator, frequently changing the setpoints of the actuators introduces considerable wear due to the high currents generated as a result of the switching. In our previous work \cite{bohn2019deep} (and indeed in other works \cite{koch2019reinforcement}) this problem is addressed through a term in the reward that discourages changing the setpoints. We now take a different approach to this problem, using the \gls{caps} method \cite{mysore2020regularizing}:

\begin{align}
    J^{\mathrm{TS}}(\pi_\vartheta) &= ||\pi_\vartheta(s_t) - \pi_\vartheta(s_{t+1})||_2 \label{eq:caps:t} \\
    J^{\mathrm{SS}}(\pi_\vartheta) &= ||\pi_\vartheta(s_t) - \pi_\vartheta(\hat{s}_t)||_2, \quad \enspace \hat{s}_t \sim \mathcal{N}(s_t, 0.01) \label{eq:caps:s}
\end{align}

This method adds two regularization terms to the loss, a temporal smoothness term \eqref{eq:caps:t} and a spatial smoothness term \eqref{eq:caps:s}. As the authors demonstrate, this method is more successful in generating controllers that yield smooth control signals compared to the action reward-term approach. Additionally, removing the action term from the reward simplifies the problem of learning the action-value function since the reward now contains fewer disparate parts, thereby accelerating learning. Instead, the gradient ascent scheme calculating the parameter updates is conditioned towards policies that are smooth in the output. Finally, we add a regularization term on the pre-activation  $\pi_\vartheta^{\mathrm{PA}}$ (that is, before applying the hyperbolic tangent in \eqref{eq:sac:pi}) of the output:

\begin{align}
    J^{\mathrm{PA}}(\pi_\vartheta) &= ||\pi_\vartheta^{\mathrm{PA}}(s_t)||_2
\end{align}

This helps in reducing the gain of the controller, especially for small errors, as it essentially tells the controller that it needs to have a strong benefit to move the actuators away from the trim-point. Additionally, we find it accelerates learning as the controller is biased towards non-aggressive control, which in conjunction with \gls{her} means the controller quickly discovers how to achieve the sparse stabilizing objective. Thus, the objective we optimize is defined as:

\begin{align}
    \begin{split}
    J(\pi_\vartheta) &= J^{\mathrm{SAC}}(\pi_\vartheta) + \lambda^{\mathrm{TS}}J^{\mathrm{TS}}(\pi_\vartheta) +  \lambda^{\mathrm{SS}}J^{\mathrm{SS}}(\pi_\vartheta) \>  + 
    \\ &\quad \> \, \lambda^{\mathrm{PA}}J^{\mathrm{PA}}(\pi_\vartheta)
    \end{split}
     \\
     \lambda^{\mathrm{TS}} &= 5\cdot10^{-2}, \enspace \lambda^{\mathrm{SS}} = 10^{-1}, \enspace \lambda^{\mathrm{PA}} = 10^{-4}
\end{align}

\subsection{UAV Model}\label{sec:method:uavmodel}
For the simulated environment, we use a model of the Skywalker X8 UAV based on previous modeling efforts~\cite{gryte2018,coates2019,winter2019,grytethesis}. The model structure is standard in the literature~\cite{Beard2012,Stevens2016} and is based on a single rigid body Newton-Euler formulation affected by forces and moments due to gravity, aerodynamics, and propulsion effects. An estimate of the inertia matrix is provided in~\cite{grytethesis} based on bifilar pendulum experiments. Results from wind-tunnel experiments are provided in~\cite{gryte2018} for aerodynamic coefficients, and in~\cite{coates2019} for the propulsion system model. This data is complemented by computational fluid dynamics (CFD) simulations from~\cite{gryte2018,winter2019}. For a more detailed description of the simulation model, see our previous work in~\cite{bohn2019deep}.

We collected a short data series to assess the quality of the dynamic model. To excite the system dynamics, we used the actuator signals from the baseline attitude controller and perturbed them with chirp signals before mapping them to the elevon deflections. The start and end frequencies of the chirp signals were $\SI{8}{\hertz}$ and $\SI{12}{\hertz}$, respectively. A dynamic mode analysis of the model indicates that this is the dominant frequency spectrum of the X8. The signal duration was 15 seconds and we used a peak-to-peak amplitude of 20 degrees. 

\begin{figure}[htb]
    \centering
    \includegraphics[width=0.5\textwidth]{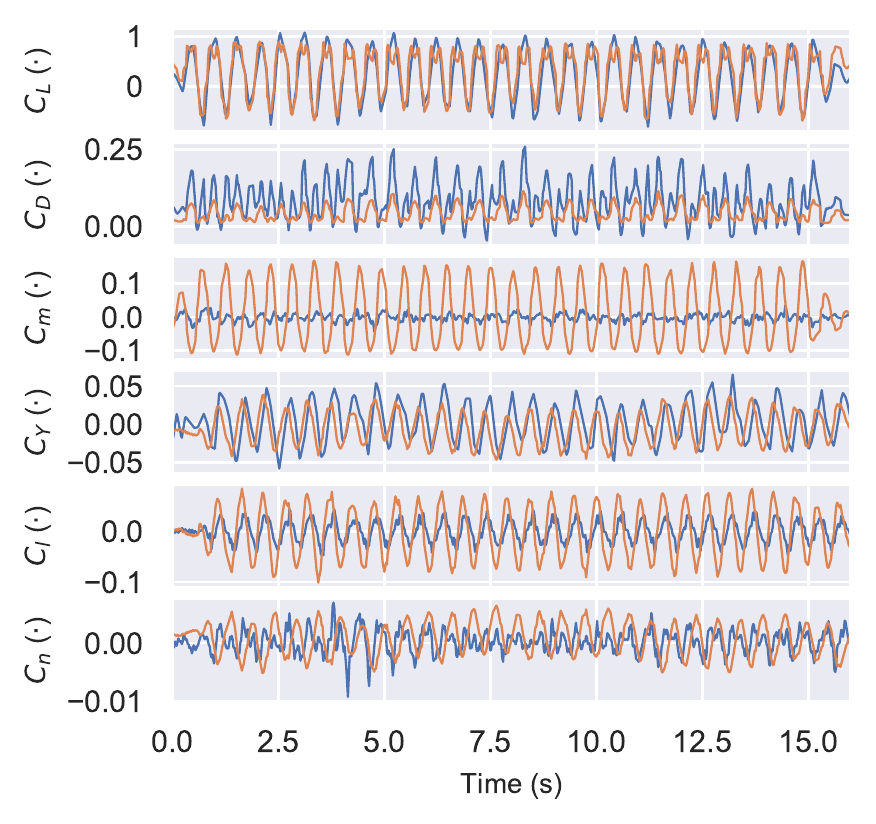}
    \caption{The aerodynamic coefficients of the UAV in a longitudinal (top three) and a lateral (bottom three) chirp signal test sequence for elevator and aileron, respectively, based on IMU data (blue) and model prediction (orange).}
    \label{fig:exp:model_valid}
\end{figure}

The aerodynamic coefficients that are calculated based on recorded sensor data and the inertia matrix of the vehicle are shown in Fig.~\ref{fig:exp:model_valid}. Following~\cite{Beard2012}, the coefficient subscript $L,D,Y,l,m,n$ denotes lift, drag, side force, roll moment, pitch moment and yaw moment, respectively. These results show that despite the modeling efforts, there are still significant discrepancies between the predicted and measured data, particularly in the pitching moment coefficient, $C_m$.

\subsection{Sim-to-Real Measures}\label{sec:method:s2r}
The main sim-to-real measure employed in the method is domain randomization. As shown in Section~\ref{sec:method:uavmodel}, there is a significant reality gap, and as such we want to avoid the \gls{rl} controller overfitting to the simulation environment. The intuition behind the domain randomization technique is that learning over a distribution of possible \gls{uav} models should robustify the controller. To this end, we assess the uncertainty in the estimate of every parameter of the \gls{uav} model and use this uncertainty to construct a probability distribution over its range of probable values (see \cite{repo} for details). The coefficients of the rate-dependent terms of the \gls{uav} model have larger sampling ranges since these are not estimated based on wind tunnel data but rather on uncertain CFD simulations \cite{gryte2018}. The sampled values are also clipped as indicated to avoid extreme unrealistic values. At the start of every episode, we sample a value for each parameter from its distribution, together constituting one realization of the \gls{uav} model.

The \gls{uav} sensor suite is subject to noise in its measurements. To model these, we first estimated the real hardware's noise characteristics, then we emulated this in the simulator environment. We model the measurement noise as an Ornstein-Uhlenbeck (OU) process \eqref{eq:measurement_noise}, which in addition to white noise gives rise to effects like measurement drift:

\begin{align}
    \hat{m}_t &= m_t + w_t, \enspace w_t \sim OU(\mu_m, \sigma_m, \theta_m), \enspace \mu_m = 0, \enspace \theta_m = 1  \nonumber  \\
    \sigma_m &= 0.005 \cdot \left[1.5, 1.5, 1.5, 1, 1, 15, 0, 0, 0, 0, 1, 1, 0, 0\right]^\top \label{eq:measurement_noise}
\end{align}

where $\mu_m, \sigma_m, \theta_m$ are the mean, variance, and rate of mean-reversion parameters of the measurement noise. Note that we do not add noise to the error and integrator states, as these are already affected by the noise in the measurement of the state that is used to calculate the error, while the previous output of the controller is by nature free of noise. The sensor suite has an update rate of 50Hz, and we therefore chose this as the control frequency as well. In the simulation environment we add exponentially distributed noise on top of the fixed control frequency in order to simulate sensor timing-variability:

\begin{align}
    \Delta_t = \Delta_0 + z_t, \enspace z_t \sim \mathrm{Exp}(\kappa), \enspace \kappa \sim \mathcal{U}(250, 1000)
\end{align}

where $\Delta_t$ is the simulation step size at step $t$, $\Delta_0 = \SI{0.02}{\second}$ is the base control frequency, and $z_t$ is exponentially distributed noise whose parameter $\kappa$ is drawn uniformly at the start of every episode.

Although not strictly sim-to-real measures, the adjustments to the \gls{rl} objective described in Section \ref{sec:method:reward} in the form of \gls{caps} and the pre-activation term also serve to improve the transferability from simulation to reality, as they encourage less aggressive lower-gain control. Another major effect present in the field is atmospheric disturbances such as wind and turbulence. We model turbulence with the Dryden turbulence model~\cite{Beard2012}, and a steady wind component whose direction and magnitude between $\SI{0}{\meter\second^{-1}}$ and $\SI{15}{\meter\second^{-1}}$ is sampled at the start of each episode. The last effect we found was highly impactful for successful transfer was the actuation delay, i.e. the time it takes before the output of the controller is applied to the system, a result which was also found in \cite{wada2021unmanned}. The simulator contains a constant actuation delay of $\SI{100}{m\second}$, while we believe this is a significant overestimation of the delay of the real system, we motivate this choice in Section~\ref{sec:method:itdev}.

\subsection{Simulator Episode Design}
The standard design of episodes for \gls{uav} control in the literature seems to be short episodes with a single constant desired attitude \cite{koch2019reinforcement,bernini2021few}. We found that having constant setpoints accelerates learning, however, the operation of the \gls{uav} in the field typically sees the navigation system continuously update the desired attitude. To align these considerations, we employ fairly long episodes of length 900 steps ($\SI{18}\second$) where setpoints are kept constant but resampled every 150 steps. To ensure sufficient diversity of the state trajectories and transitions used to update the parameters of the \gls{rl} controller, we sample initial conditions as shown in Table~\ref{tab:ep_init}. Considering that the main objective of the simulation environment is to ready the controller for the field, we sample initial conditions mostly from the linear region of the model, as this is where the \gls{uav} model is assumed to have the most validity. Note that while the range of initial conditions is somewhat limited, there is nothing stopping the controller from exploring the full state space. Furthermore, since the initial states of the actuators are also randomized the sampled initial conditions could cause instability, such that the controller must learn to recover. 

\begin{table}
    \vspace*{0.13cm} 
    \centering
    \caption{Initial conditions for the episodes are uniformly sampled from the indicated ranges.}
    \ra{1.2}
    \begin{tabular}{@{}lrllrl@{}} \toprule
    Variable & Range & Unit & Variable & Range & Unit \\ \midrule
    $\phi$ & -40, 40 & degrees & $\phi_r$ & -60, 60 & degrees \\
    $\theta$ & -15, 15 & degrees & $\theta_r$ & -25, 20 & degrees \\
    $V_a$ & 13, 26 & m/s & $\alpha$ & -8, 8 & degrees \\
    $\omega$ & -60, 60 & degrees & $\beta$ & -10, 10 & degrees \\
    $\delta_{r, l}$ & -30, 30 & degrees \\
    \bottomrule
    \end{tabular}
    \label{tab:ep_init}
\end{table}

\subsection{Reinforcement Learning Algorithm}
To develop the \gls{rl} controller we use the \gls{sac} algorithm and augment the collected data using \gls{her} \cite{andrychowicz2017hindsight}, see \cite{repo} for hyperparameters and code. We chose the \gls{sac} algorithm because it is off-policy, and therefore has comparatively high data efficiency among \gls{rl} methods, and furthermore the policy is explicitly trained to handle perturbations from the inherent randomness, which tends to yield more robust policies that transfer better than non-entropy-regularized algorithms. Note that we employ the technique of initializing the replay buffer of the algorithm with 5k data samples (corresponding to 100 seconds of flight at 50Hz), which is a common technique in \gls{rl} to help the policy with the initial exploration phase. This data is entirely independent of the learning controller being trained and is obtained by uniformly sampling random actions from the set of possible actions in the simulator environment. This data could also stem from other sources such as historical data gathered by a human pilot or another controller, which might be more suitable when performing exploration exclusively in the field. Since this data is independent of the learning controller, we do not count it towards the data requirement of our method and do not include it in the learning curve graphs.

\subsection{Experimental Platform} \label{sec:method:platform}
Our custom avionics stack is based on the low-level control architecture developed at the NTNU UAV-lab~\cite{coates2022}. It consists of a Cube Orange flight controller running the (industry standard) ArduPlane open-source autopilot~\cite{ardupilot}, and a Raspberry Pi 4 running the DUNE Uniform Navigation Environment~\cite{lstsdune}. During experiments, the total flight weight of the Skywalker X8 is \SI{3.8}{\kilogram}. 

The \gls{rl} controller is implemented as a DUNE task in C++ with the \gls{nn} implemented in TensorFlow. Sensor data and state estimates from ArduPlane are sent through a serial connection to the Raspberry Pi, providing all necessary data for the \gls{rl} controller. The \gls{nn} controller output is converted to PWM duty cycle and sent to the elevon servos using a PCA9685 servo driver, interfaced through I2C from the Raspberry Pi. A PWM multiplexer supports switching between the \gls{rl} controller output and ArduPlane. This enables the pilot to always take control during testing, either through manual controls, or through ArduPlane's standard autopilot. This switching mechanism enables us to safely engage the \gls{rl} controller in flight, while takeoff and landing are performed by the pilot operating our tried and tested avionics stack~\cite{zolich2015}.

\section{Experimental Results}\label{sec:experiments}

This section presents the main experimental results, collected during two days of flight experiments at Agdenes Airfield, Breivika, Norway in September 2021. During the first day, we enjoyed calm weather and perfect flight conditions, with a mean wind speed (as estimated by ArduPlane's Kalman Filter) of less than $\SI{4}{\meter\per\second}$. The second day of flight tests, however, presented challenging weather conditions, with frequent gusts and a mean wind speed of approximately $\SI{12.5}{\meter\per\second}$ (70\% of the Skywalker X8's cruise speed of $\SI{18}{\meter\per\second}$).

We present three types of data, differing mainly by how roll and pitch angle references are provided: 
\begin{enumerate}
    \item References are given by the pilot, mimicking ArduPlane's \gls{fbwa} mode (Section~\ref{sec:exp:fbwa}).
    \item References are provided by ArduPlane's guidance system, which is set to track a rectangular waypoint pattern (Section~\ref{sec:exp:auto}).
    \item References are set by a predefined, automated series of steps (Section~\ref{sec:exp:step}). Similar maneuvers are also performed with an implementation of the ArduPlane \gls{pid} attitude controller, with the response compared to that of the \gls{rl} controller.
\end{enumerate}
In contrast to the training phase, where the throttle actuator used to control airspeed is operated by a PI controller (see~\cite{bohn2019deep} for details), the throttle is either under manual control by the pilot (\gls{fbwa}) or controlled by ArduPlane (auto/steps). In figures presenting flight results, the dashed orange line corresponds to state reference, while in the elevon plots, the blue and orange lines correspond to the right and left elevon, respectively, with minimum and maximum deflections of $-30\degree$ and $+30\degree$.  

\subsection{FBWA Mode} \label{sec:exp:fbwa}
\begin{figure}[htb]
    \centering
    \includegraphics{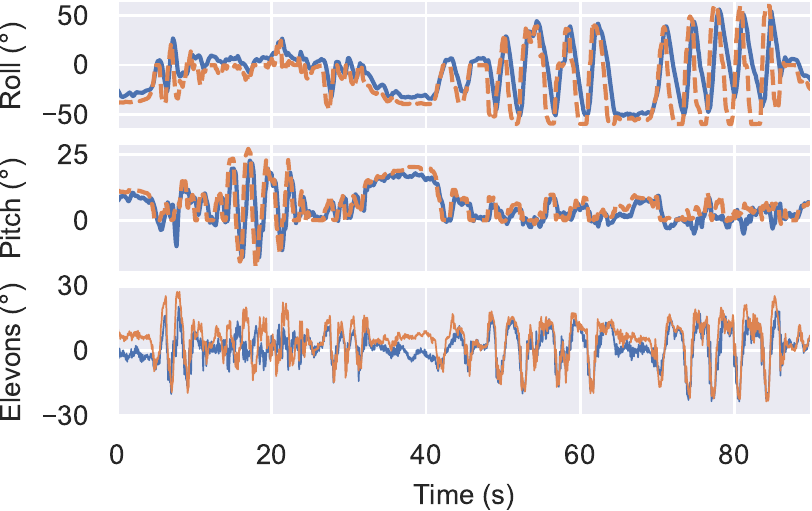}
    \caption{\Gls{fbwa} mode using references (dashed orange line) from the pilot, showing the attitude states and right (blue) and left (orange) elevon signals.}
    \label{fig:exp:pilot}
\end{figure}

Fig.~\ref{fig:exp:pilot} shows an excerpt from the flight experiments where a human pilot decides the desired attitude of the \gls{uav}. The \gls{rl} controller is able to closely track the desired attitude even for the most difficult and aggressive maneuvers, while producing smooth outputs for the actuators. We do however note a consistent steady-state error. Towards the end of the maneuver, we observe that the roll response is non-symmetric, that is, rolling to the left (towards negative roll angles) is slower than rolling in the opposite direction. We investigate and discuss this matter, as well as the steady-state error, in Section~\ref{sec:discussion}.

\subsection{Auto mode} \label{sec:exp:auto}

\begin{figure}[h]
    \centering
    \subfloat{\includegraphics[]{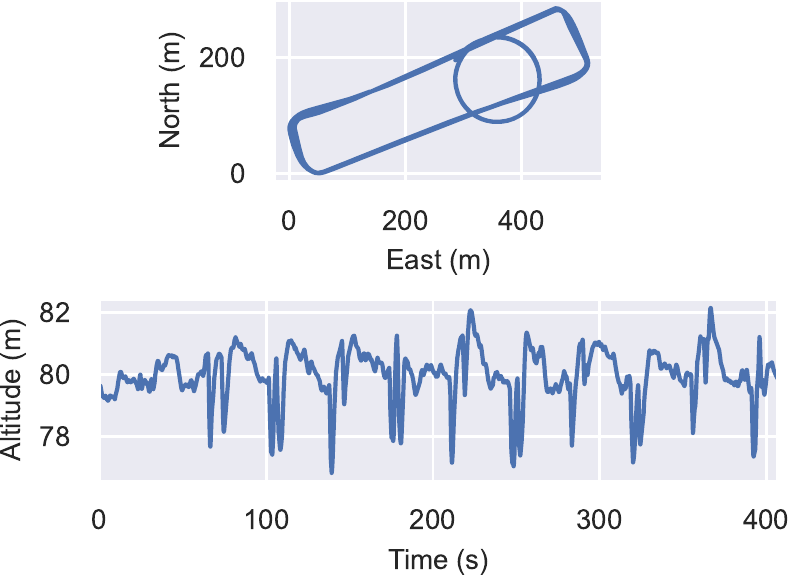}}
    \caption{Position plot showing how the \gls{rl} controller is able to take references from ArduPlane's guidance system in Auto mode, and effectively follow prescribed paths. First, a loiter, then a square waypoint pattern.}
    \label{fig:exp:auto}
\end{figure}

Fig.~\ref{fig:exp:auto} shows the results for the \gls{rl} controller operating with references provided from the ArduPlane guidance system, set to track a square waypoint pattern. Before tracking the square, the \gls{uav} loiters in a circular pattern for a while. Despite some steady-state offset, especially for the roll angle error, the \gls{uav} successfully completes the specified mission. This is because the outer-loop guidance controller can compensate for this error, and still achieve convergence when faced with disturbances such as wind and offset in inner-loop control. This is similar to how a pilot supplying manual references would offset the references to keep the intended course.

During turns, a certain altitude drop is seen from the right part of Fig.~\ref{fig:exp:auto}. This is caused by the aggressive turn radius accompanied by drops in the pitch angle. This effect can be reduced by tuning the guidance system to be less aggressive (e.g. by increasing the turn radius) or reducing the maximum allowable roll angle setpoint, which is set to be 55 degrees. A similar drop in pitch angle is also seen when using the default ArduPlane attitude controller. 

\subsection{Step sequences and Comparison with ArduPlane PIDs} \label{sec:exp:step}
\begin{figure*}
    \centering
    \subfloat{\includegraphics[]{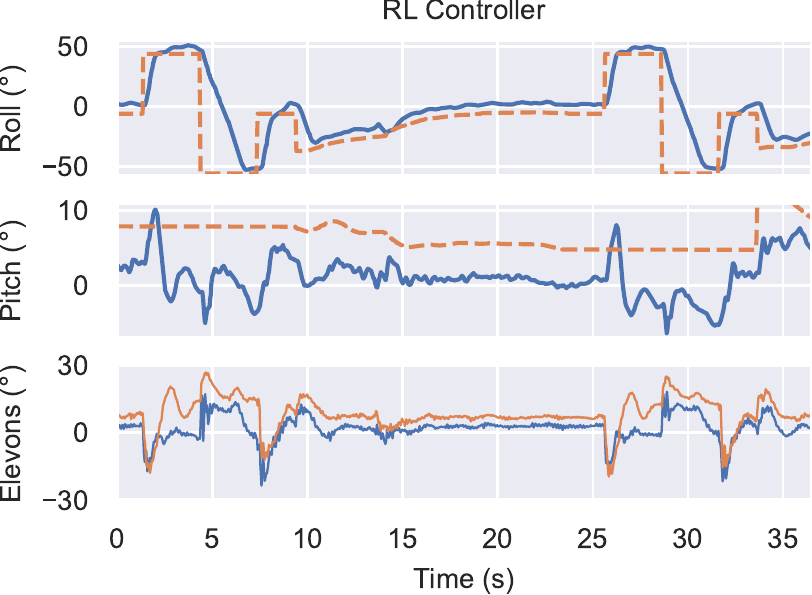}}~
    \subfloat{\includegraphics[]{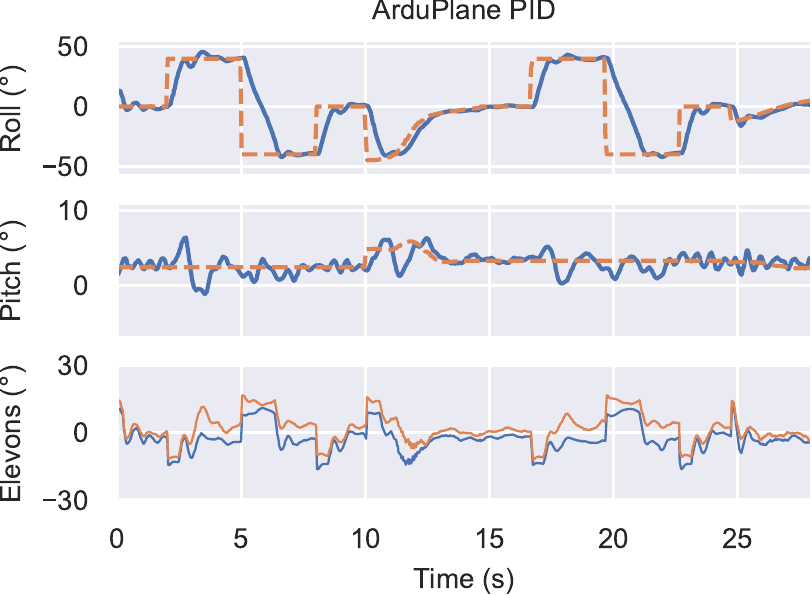}}
    
    \subfloat{\includegraphics[]{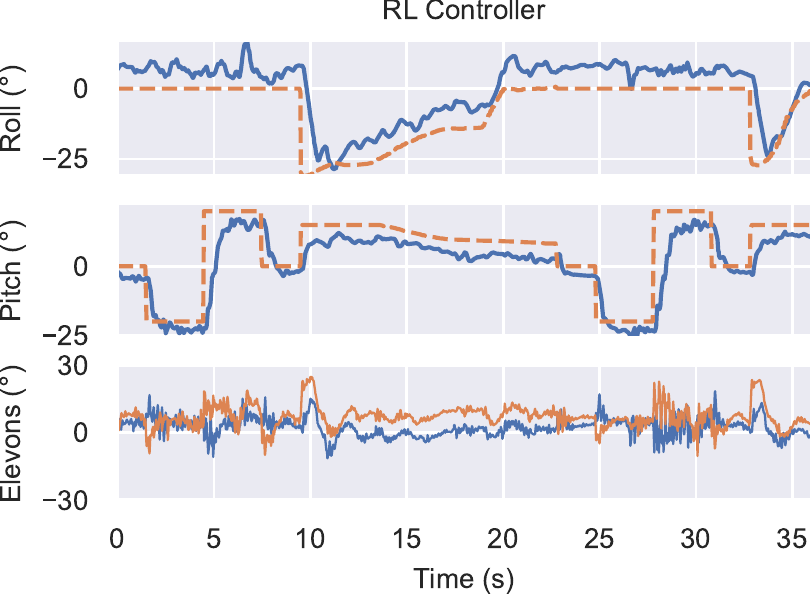}}~
    \subfloat{\includegraphics[]{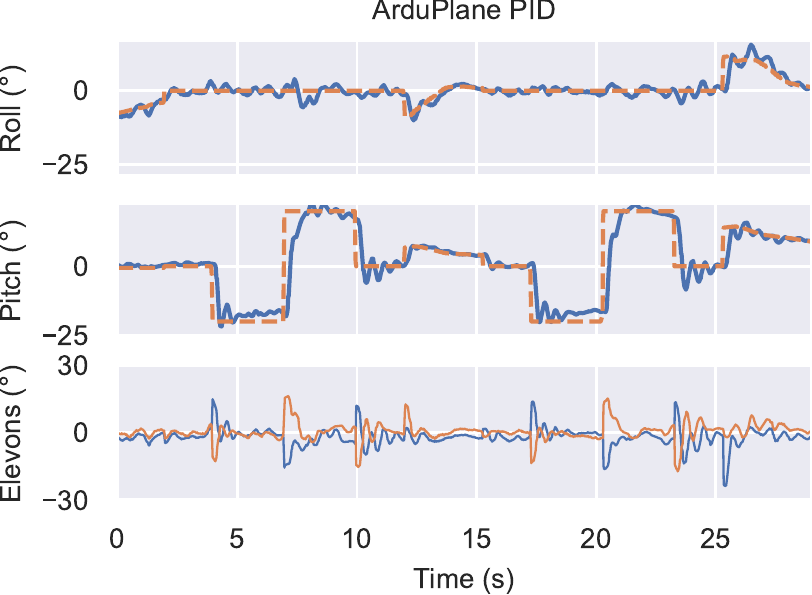}}
    \caption{Comparison between the \gls{rl} controller and the ArduPlane PID controller for steps in the roll (top) and pitch (bottom) references (dashed orange line). }
    \label{fig:exp:pid_com_Steps}
\end{figure*}

Step responses for the \gls{rl} controller, as well as the ArduPlane PID controller, are displayed in Fig.~\ref{fig:exp:pid_com_Steps}. The \gls{rl} controller shows comparable transient performance to that of ArduPlane, the main difference being the steady-state error of the \gls{rl} controller. Additionally, the pitch response of the \gls{pid} controller is slightly more aggressive. However, this could be changed by tuning the controller. 

The control signals generated by the \gls{rl} controller are relatively smooth and well-behaved but include some high-frequency components not seen in the \gls{pid} response. Apart from that, the control input looks qualitatively similar, with similar magnitude. For a quantitative comparison we employ the smoothness metric defined in \cite{mysore2020regularizing} which jointly considers the amplitudes and frequencies of the control signals:

\begin{align}
    Sm = \frac{2}{n f_s} \sum_{i=1}^n M_i f_i \label{eq:sm_metric}
\end{align}

where $M_i$ is the amplitude of the $i'\mathrm{th}$ frequency component $f_i$, and $f_s$ is the sampling frequency. On this metric the \gls{pid} measures at $6.20\cdot 10^{-4}, 6.30 \cdot 10^{-4}$ for the roll and pitch maneuvers in Fig.~\ref{fig:exp:pid_com_Steps}, respectively, while \gls{rl} measures $2\%$ and $44\%$ higher at $6.30 \cdot 10^{-4}, 9.05 \cdot 10^{-4}$. This metric shows that \gls{rl} has comparable smoothness in its output with the \gls{pid} controller, but also indicates the higher frequency components of the \gls{rl} controller's output in Fig.~\ref{fig:exp:pid_com_Steps}. It is not clear why there is such a discrepancy between the two maneuvers for the \gls{rl} controller, but this data is as mentioned subjected to considerable turbulence and wind, and could therefore be caused by transient gusts.

While the former results were gathered on a calm day with virtually no wind, these maneuvers are executed in harsh wind conditions on day two. The \gls{uav} also suffered structural damage (not while under \gls{rl} control) after collecting the \gls{pid} data, before redoing the experiments with \gls{rl}. The vehicle had to be repaired with a new wing and duct tape, causing a change in the \gls{uav}'s trim point. Thus, the presented results demonstrate the \gls{rl} controller's robustness towards model mismatch and varying wind conditions, including heavy gusts.

To achieve a fair comparison, the ArduPlane PID is implemented in the same software stack and ran with the same hardware architecture as the \gls{rl} controller (described in Section~\ref{sec:method:platform}). In particular, this means that any increased signal latency introduced in our setup does not affect the comparison.

\section{Discussion}\label{sec:discussion}
The experimental results of Section~\ref{sec:experiments} show that the \gls{rl} controller performs well compared to a state-of-the-art open-source autopilot, and is robust to disturbances caused by harsh wind conditions. The control performance of the \gls{rl} controller across the various flight modes speaks to its ability to generalize further than just the maneuvers encountered during training. In particular, no guidance controller was present during training.

Despite the promising results, there is room for improvement. In this section, we further discuss how performance can be improved, the iterative development process, and training, and we perform a linear analysis to gain further insight into the behavior of the \gls{rl} controller.

As noted in Section~\ref{sec:exp:fbwa}, the roll response of the \gls{rl} controller is non-symmetric, meaning that rolling towards the left wing is slower than rolling to the right. This is supported by the pilot's qualitative assessment during flight. We found that this effect was caused by an overestimation of the simulated propeller torque, which was found to be less prominent on the physical \gls{uav} than expected, presumably due to the mechanical mounting of the propeller. This causes a bias in the \gls{rl} controller, which has learned to counteract the propeller torque. To remedy this, we trained a new \gls{rl} controller in a simulation model without propeller torque, which was briefly tested in the field to verify our hypothesis. The new controller exhibited a more symmetric roll response, as expected.

\subsection{Steady-State Errors}\label{sec:disc:int}
We experimented with several techniques in order to address the steady-state error of the \gls{rl} controller observed in flight experiments: pure integrator (no decay), higher decay factor (e.g. 0.999), having integration separate from the \gls{nn} controller with learned integration gains, shaped rewards, and training with input disturbances. Whereas some of these measures reduced the steady-state error to some degree, none were successful in entirely eliminating it.

We note that there is no consistent steady-state error in the simulator in the same way we observe in field experiments, i.e. consistently over or under the reference with a consistent magnitude. The controller has learned to use integral action to reduce steady-state error from disturbances in the simulator, but not in a way that transfers to the field. This could be because the controller is overfitting to the simulator, thus the larger tracking errors in the field combined with the hyperbolic tangent saturating functions of the \gls{nn} causes the integrator states' effect on the output to saturate prematurely.  

\begin{figure}
    \centering
    \includegraphics[]{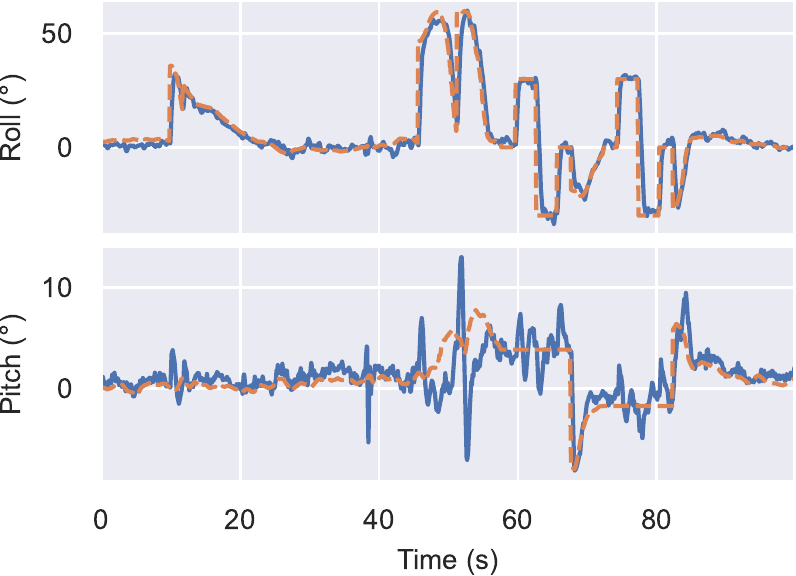}
    \caption{Response of the \gls{rl} controller where the steady-state error has been estimated and references (dashed orange line) adjusted to compensate.}
    \label{fig:disc:trim_ref}
\end{figure}

An effective way to address this problem is to estimate the steady-state error and then add the estimated value to the references provided to the \gls{rl} controller, as was done in the flight experiment shown in Fig.~\ref{fig:disc:trim_ref}. As can be seen, this simple technique can fully compensate for the steady-state error and may also be automated using an integrator in an outer loop to estimate the steady-state error \cite{martinsen2018end}. Furthermore, there are compelling arguments for not having integral action in the inner-loop attitude controller, as adding integral action to the controller necessarily reduces the phase margins and the achievable bandwidth \cite{Beard2012}.

\subsection{Oscillations: Illustration of Iterative Development}\label{sec:method:itdev}
\begin{figure}
    \centering
    \subfloat{\includegraphics[]{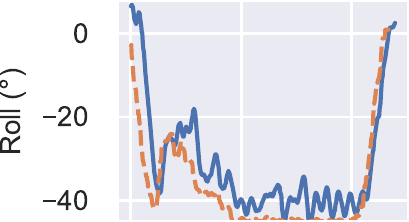}}~
    \subfloat{\includegraphics[]{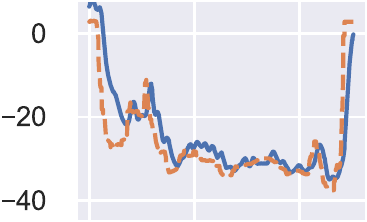}}
    
    \subfloat{\includegraphics[]{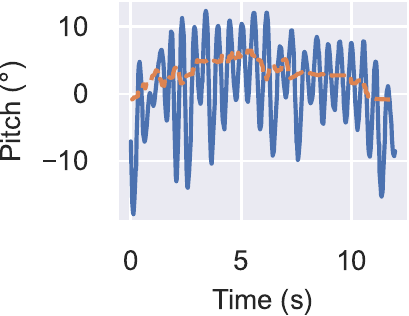}}~
    \subfloat{\includegraphics[]{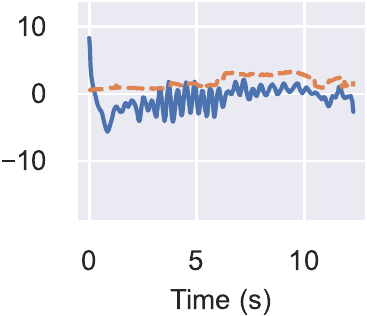}}
    \caption{Oscillatory attitude response of initial flight experiments, for one controller trained with original (left) and one with reduced (right) $C_{m_q}$.}
    \label{fig:itdev:cmq}
\end{figure}

Initial field experiments were characterized by excessive oscillations in the attitude response of the \gls{uav}, especially in pitch, necessitating halving the \gls{rl} controller's outputs in order to keep the aircraft airborne. These oscillations were not present in the simulator, as such, we suspected that this was (at least in part) caused by the simulator overestimating the natural damping present in the aircraft. Therefore, we reduced the $C_{m_q}$ (pitch damping) parameter by a factor of 10. While this reduced the oscillations somewhat, there were still significant oscillations in the response, see Fig.~\ref{fig:itdev:cmq}.

We estimated a typical actuation latency for the system of $\SI{10}{\meter\second^{-1}}$. In sim-to-sim experiments, where we raised the latency of the control system during the exploitation phase of a controller trained with $\SI{10}{\meter\second^{-1}}$ latency, we observed similar oscillatory responses as in the flight experiments and noted its relationship with increasing latency. We then trained an \gls{rl} controller where the latency was set to $\SI{100}{\meter\second^{-1}}$ during the learning phase. This controller trained with higher latency almost entirely eliminated the oscillations to the levels shown in the field experiment figures. Favoring robust controller design, we increased the base latency of the simulation environment to $\SI{100}{\meter\second^{-1}}$, even though we believe that this is a significant overestimation of the true latency of the real system.

\subsection{Linear Analysis}\label{sec:experiment:analysis}
\begin{figure}[ht]
    \centering
    \includegraphics[width=0.5\textwidth]{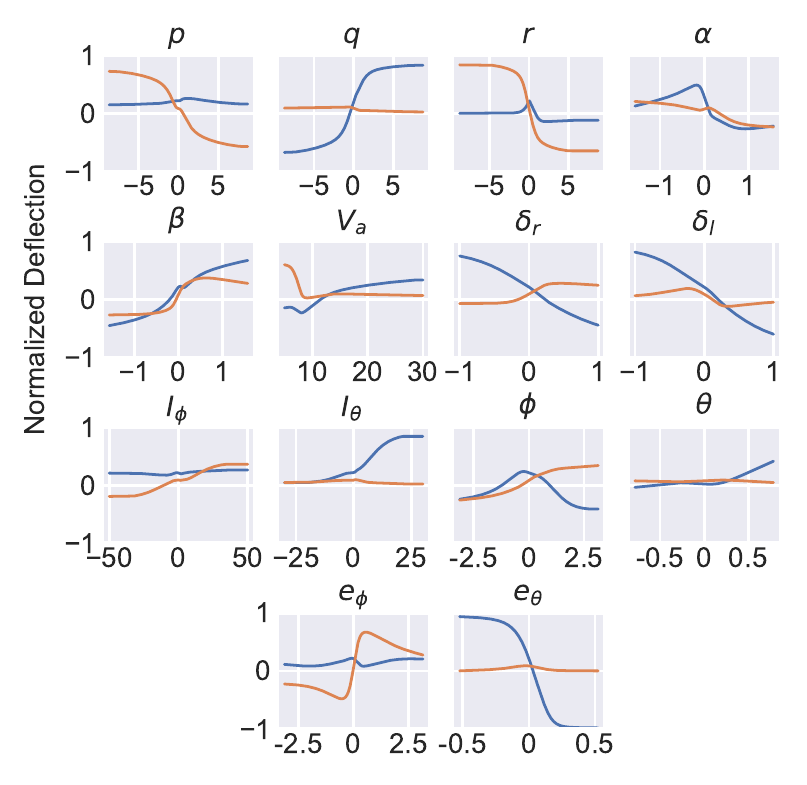}
    \caption{Open-loop level-flight response of the RL controller when perturbing one input at a time. The x-axis is in the units of the corresponding state. The lines are elevon outputs mapped to aileron (orange) and elevator (blue).}
    \label{fig:response}
\end{figure}

In order to better understand how the \gls{rl} attitude controller operates, we analyze its sensitivity to the input variables. In Fig.~\ref{fig:response} we have plotted the open-loop response of the controller as a function of a single perturbed input. The rest of the state vector is kept constant at the steady-flight value, i.e. zero for all variables except the airspeed $V_a$ which is set to the cruising speed of $\SI{18}{\meter\per\second}$, and the angle-of-attack $\alpha$ and pitch angle $\theta$, which are kept at the trim values necessary to generate lift for level flight, while the input values for previous time steps are kept constant in the time dimension. To be able to compare the results with ArduPlane, we translate the elevon outputs into virtual elevator and aileron commands using the inverse of~\eqref{eq:linmap1}-\eqref{eq:linmap2}. The figures and tables are presented in terms of these virtual commands, which also have a more intuitive and straightforward effect on the roll and pitch angles.

The saturating effect of the hyperbolic tangent nonlinearity on the \gls{rl} controller is distinctly present in the responses. This is a desired effect as we know that any input should have a bounded effect on the output, which gives robustness towards possible measurement errors or misalignment of the dynamics of the simulation and real environments. The controller makes use of all its inputs, with the previous outputs of the controller having the most significance for the current output (the typical values for most states in Fig. \ref{fig:response} are close to the level-flight value in the center and will thus use a limited range of the response curve, while the previous output of the controller frequently employs the full range). This makes sense as the controller is conditioned towards smooth outputs, as described in Section~\ref{sec:method:reward}, which means that a reasonable initial guess of any action is to be similar to the previous action. Moreover, the fact that the previous output (left and right elevons) do not have a symmetric effect on the subsequent output (there is no mechanism enforcing symmetry in the learning controller) could be a motivating factor to instead employ (virtual) elevator and aileron as outputs of the \gls{rl} controller.

Since the control authority of the actuators increases with airspeed, we investigated if the RL controller has learned to scale its outputs depending on airspeed, an effect that is included in the ArduPlane controller. With no further documentation (due to space constraints), we state that this is not the case. However, it has learned to bias the response, essentially shifting the curves in Fig.~\ref{fig:response} up, as airspeed increases in order to compensate for the change in trim-point with airspeed.

To estimate sensitivities wrt. an input we take a linear approximation of its response curve by using the slope of the tangent line at level-flight conditions. The result is shown in Table~\ref{tab:gains}, and compared to the ArduPlane \gls{pid} controller gains (see Appendix \ref{sec:app:arduplane}). The \gls{rl} controller is noticeably more aggressive in the pitch error, while simultaneously introducing more damping through the angular velocity component $q$. This is evident in Fig.~\ref{fig:exp:pid_com_Steps} where the \gls{rl} controller exhibits less oscillation in the pitch response. The estimated gains for the integrator states in Table~\ref{tab:gains} are not representative of the response curves for these states, as the response curve exhibits cubic characteristics with a small opposing region around the level-flight value. Thus, for these states, a linear approximation over a larger region would be more descriptive. Overall, the gains of the \gls{rl} controller are similar to those of the \gls{pid} controller, which increases the trust in the \gls{rl} controller. On the other hand, the dynamic aspect caused by integral states and data from previous time steps increases the complexity of the analysis and thus limits the conclusions that can be drawn from it.

\begin{table}
    \centering
    \caption{Linearly approximated gains at level-flight, where each input is perturbed in isolation.}
    \ra{1.2}
    \begin{tabular}{@{}lrrrrrr@{}} \toprule
     \multirow{2}{*}{Controller} & \multirow{2}{*}{$\cfrac{\partial\delta_a}{\partial e_\phi}$} & \multirow{2}{*}{$\cfrac{\partial\delta_e}{\partial e_\theta}$} & \multirow{2}{*}{$\cfrac{\partial\delta_a}{\partial I_\phi}$} & \multirow{2}{*}{$\cfrac{\partial\delta_e}{\partial I_\theta}$} & \multirow{2}{*}{$\cfrac{\partial\delta_a}{\partial p}$} & \multirow{2}{*}{$\cfrac{\partial\delta_e}{\partial q}$}
     \\ \\ \midrule
    RL & 1.268 & -3.320 & -0.005 & 0.006 & -0.008 & 0.223 \\
    PID & 1.630 & -1.081 & 0.052 & -0.052 & -0.024 & 0.031 \\
    \bottomrule
    \end{tabular}
    \label{tab:gains}
\end{table}

\subsection{Data Requirements}

\begin{figure}[ht]
    \centering
    \includegraphics[width=0.5\textwidth]{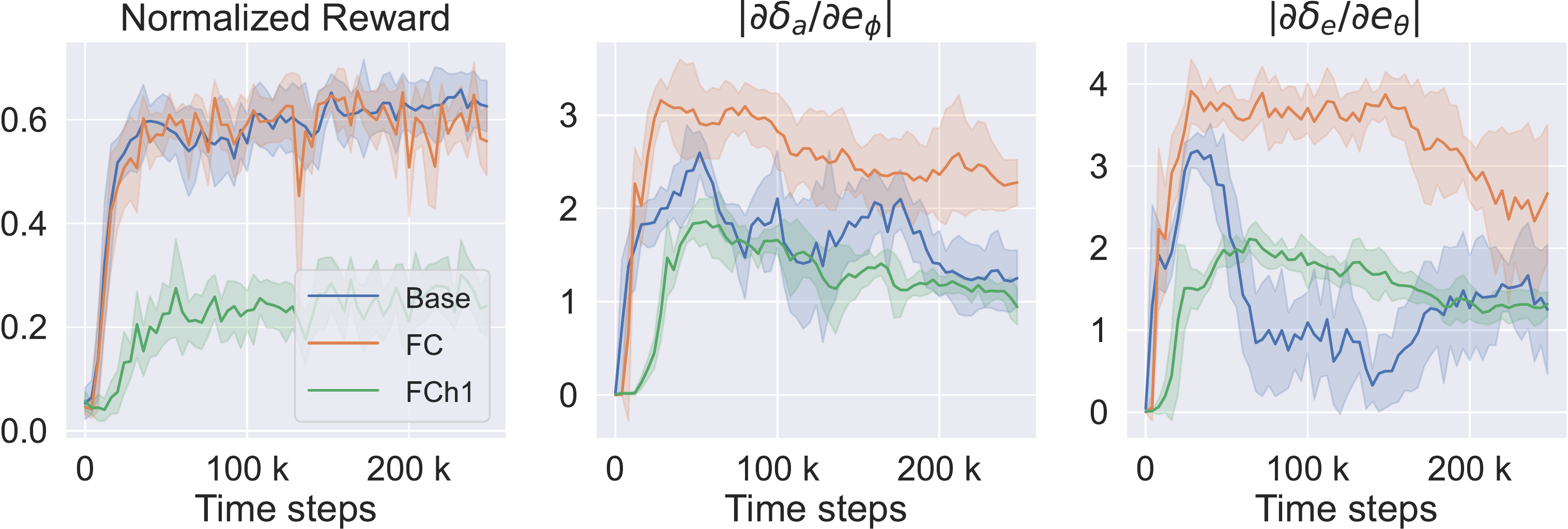}
    \caption{The learning phase of the proposed \gls{rl} controller, showing normalized mean episode reward and error-proportional gains. The solid line represents a rolling average mean value while the shaded region represents one standard deviation over three randomly seeded controllers. Base refers to the method as presented in Section~\ref{sec:method}, the \gls{fc} version replaces the convolutional layer with \pgls{fc} layer, and the h1 version has no history in the input.}
    \label{fig:training}
\end{figure}

\begin{figure}[htb]
    \centering
    \includegraphics{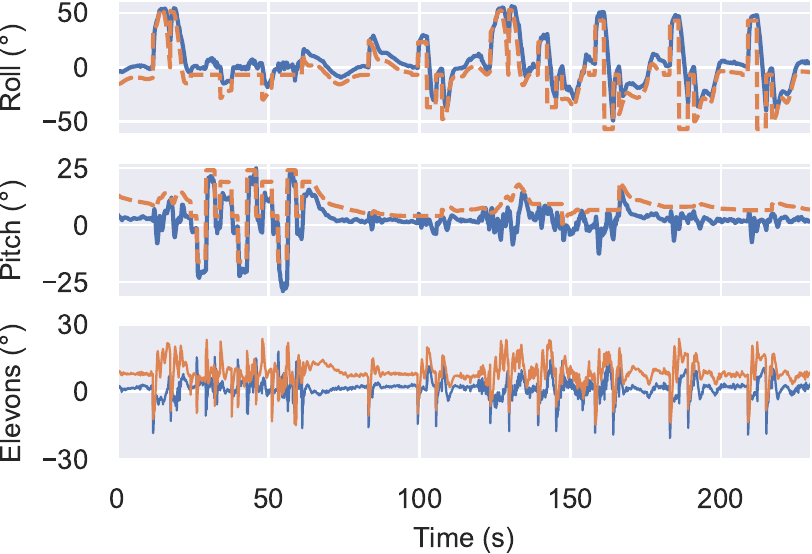}
    \caption{An \gls{rl} controller that has trained for only 3 minutes of real-time flight in the simulation environment is airworthy, able to track references (dashed orange line) reasonably well.}
    \label{fig:exp:es}
\end{figure}

In this section, we attempt to quantify the data efficiency of our method. We start by presenting the learning phase of the controller in the simulation environment (Fig.~\ref{fig:training}), demonstrating that our method produces proficient controllers with a number of data samples on the order of 10s of thousands, and then demonstrate that the learning controllers are flight-worthy in the field after experiencing only three minutes of real-time flight data (Fig.~\ref{fig:exp:es}). It is difficult to compare this result directly to the existing literature, for reasons outlined in the literature review of Section \ref{sec:introduction}, that is, no other reported work study the full attitude control problem of fixed-wing \glspl{uav} using \gls{rl}. Other works study either a limited version of the attitude control problem, or consider other aircraft designs (e.g. quadcopters) whose dynamics are more linear and controllable than the dynamics of a fixed-wing \gls{uav}. Nonetheless, works in the existing literature report a data requirement on the order of 100s of thousands or millions of data samples, meaning our result presents a significant improvement in terms of data efficiency. Further, we would argue that this improvement in data efficiency is significant because it suggests that learning the controller entirely in the real world (i.e. no simulation required) is possible. Collecting data for a few minutes of flight seems reasonably achievable through some form of safe exploration or guided learning where, for instance, the learning controller is monitored by a pilot or other proficient controller that can assume control should dangerous situations arise.

The evolution of the learning phase for the \gls{rl} controller as a function of time steps is shown in Fig.  \ref{fig:training}. For every version in Fig.~\ref{fig:training}, we train three controllers each with a different initial random seed and average the results over the controllers. The rewards are normalized so that 1 corresponds to attaining the maximum reward as defined in \eqref{eq:method:reward} at every step (although this is not physically achievable) and 0 corresponds to obtaining no rewards at all. Base refers to the \gls{rl} controller as presented in Section~\ref{sec:method} that was used in the field experiments. To assess the contribution of the convolutional input layer, we trained one version where the input layer is replaced with \pgls{fc} layer, and further to test the importance of the history of states in the state vector we train one model with \pgls{fc} input layer and with $h = 1$ (labeled FCh1). 

The base version learns fast, reaching convergent performance after around 40k time steps. This corresponds to about 13 minutes of real-time flight or 16 minutes of wall-clock training time when trained entirely on an i7-9700 Intel CPU with a single data-generating-agent in the simulator\footnote{Note that the wall-clock training time can essentially be made arbitrarily short with parallelization of more agents and more powerful computing hardware}. We verified in the field that the \gls{rl} controller is flightworthy long before this point: The controller shown in Figure~\ref{fig:exp:es} exhibits decent and stable performance (although unable to follow the most aggressive maneuvers) after only 10k time steps of training in the simulator (corresponding to three minutes of real-time flight). We conjecture that the data efficiency of our method limits overfitting to the simulation environment and therefore transfers better to the field, although this statement requires further evidence such as a bench of learning controllers with varying time in simulation to verify.

We find that our training method is stable in the sense that the performance differences between controllers with different seeds are small, within a few percent. The \gls{fc}h1 version without state history is never able to learn to consistently stabilize the \gls{uav} at the desired attitude in the time frames we considered. The \gls{fc} version with state history on the other hand achieves comparable rewards to the base version, showing the importance of history in the \gls{rl} input state. The base version reaches peak performance slightly faster than the \gls{fc} version, and further its proportional gains are considerably lower. This is also evidenced by the smoothness metric \eqref{eq:sm_metric} for which the base version scores 50\% lower than the \gls{fc} version. The lower gains and the smoothness metric indicate that the convolutional input layer provides a superior ability to predict the system response and thus provide smoother responses in attitude and control signals, while the \gls{fc} version is more reactive and oscillatory.

\section{Conclusion}\label{sec:conclusion}
This paper has presented a data-efficient method for learning attitude controllers for fixed wing \glspl{uav} using \gls{rl}. The learning controller is able to operate directly on the nonlinear dynamics, and therefore could extend the flight envelope and capabilities of autopilots. The high data efficiency of the presented method facilitates transfer to control of the real \gls{uav} by limiting overfitting to the simulated model. We demonstrate that the learned controller has comparable performance to the existing state-of-the-art ArduPlane \gls{pid} autopilot, and is capable of tracking prescribed paths from a guidance system while generating smooth actuation signals and attitude responses. Key factors behind the success of the method were robustifying the controller through increasing its phase margins by learning with significant actuation delay and diversifying the simulated dynamics, as well as incentivizing non-aggressive control through sparse rewards and additional objective terms enforcing temporal and spatial smoothness in the controller outputs.

Further work in this direction should investigate if RL solutions to more complex flight control problems also transfer well to the field, e.g. deep-stall landings or end-to-end path following. The problem of limited integral action should also be further investigated. Moreover, learning from real data, be it historical or generated online by the learning controller, is an intriguing research direction that would alleviate the need for accurate nonlinear models. The data efficiency of our method shows that this should in fact be feasible.
\section*{Acknowledgment}
The authors would like to thank Pål Kvaløy, Kristoffer Gryte and Martin Sollie for their crucial role in planning and executing field experiments, and fruitful discussions in general about all things \gls{uav}.

\bibliographystyle{IEEEtran}
\bibliography{main}

\begin{appendices}
\section{The ArduPlane Attitude Controller} \label{sec:app:arduplane}
This section presents the main equations used for attitude control in ArduPlane~\cite{ardupilot}, which is a state-of-the-art open-source autopilot for fixed-wing UAVs. This controller is used as a baseline to compare the RL controller against and to support the discussion in Section~\ref{sec:experiment:analysis}. The equations are based on ArduPlane, Release 4.0.9, which is the most recent stable release (as of August 2021).

The ArduPlane attitude controller consists of two cascaded single-input-single-output (SISO) feedback loops. The elevator controls pitch angle, while the ailerons are used for roll control. The outer loop consists of proportional controllers, where desired roll and pitch rates $p_r,q_r \in \mathbb{R}$ are calculated according to
\begin{align}
    p_r &= k_\phi \left(\phi_r - \phi  \right) \\
    q_r &= k_\theta \left(\theta_r - \theta \right) + q_{ct} ,
\end{align}
where $k_\phi,k_\theta > 0$ and $q_{ct}$ is the pitch rate offset needed to maintain a constant pitch angle during coordinated turns, given by
\begin{equation}
    q_{ct} = \sin(\phi) \cos(\theta) \frac{g}{V_a} \tan(\phi) .
\end{equation}
The rate setpoints are inputs to the inner loop, which consists of proportional-integral (PI) controllers with feedforward action:
\begin{align}
    \delta_a &= k_{p,p} \nu^2 \left( p_r - p \right) + \int_0^t k_{i,p} \nu^2\left( p_r - p \right) d \tau + k_{ff,p} \nu p_r \\
    \delta_e &= -k_{p,q} \nu^2 \left( q_r - q \right) - \int_0^t k_{i,q} \nu^2 \left( q_r - q \right) d \tau - k_{ff,q} \nu q_r ,
\end{align}
where $k_{p,*}$, $k_{k_i,*}$ and $k_{ff,*}$ are proportional, integral and feedforward gains, respectively. The variable $\nu = V^*/V_a$, where $V^*$ is some constant reference airspeed, provides airspeed scaling of the controller parameters, accounting for the fact that larger airspeeds give greater aerodynamic control authority. The negative sign in the control law for $\delta_e$ is introduced to account for the convention that positive elevator deflections yield a negative pitch moment~\cite{Beard2012}.

For UAVs equipped with a rudder, additional control loops utilize the extra control surface for turn coordination. However, as the Skywalker X8 considered in this paper is rudderless, this part of the control algorithm is not relevant here.

For an elevon plane like the Skywalker X8, the aileron and elevator deflection angles are virtual control signals that are mapped to elevon control actions using the linear map
\begin{align}
    \delta_l &= \delta_e + \delta_a  \label{eq:linmap1}\\
    \delta_r &= \delta_e - \delta_a . \label{eq:linmap2}
\end{align}

By assuming a constant airspeed $V_a = \SI{18}{\meter\per\second}$ and inserting parameters used for the Skywalker X8 UAV at the NTNU UAV-lab, we get the following sensitivities for the elevator and aileron control signals:
\begin{align}
    \left. \frac{\partial \delta_e }{\partial e_\theta} \right\rvert_{\theta=\phi=0} &= -1.0813 \quad &\frac{\partial \delta_a }{\partial e_\phi} &= 1.6299\\
    \left. \frac{\partial \delta_e }{\partial q} \right\rvert_{\theta=\phi=0} &=  0.0312 \quad & \frac{\partial \delta_a }{\partial p} &=  -0.0243\\
    \left. \frac{\partial \delta_e }{\partial I_\theta}\right\rvert_{\theta=\phi=0} &= -0.0521 \quad & \frac{\partial \delta_a }{\partial I_\phi} &= 0.0521\\
    \left. \frac{\partial \delta_e }{\partial \theta}\right\rvert_{\theta=\phi=0} &= 0.0104 \quad & \frac{\partial \delta_a }{\partial \phi} &= -0.0104 ,
\end{align}
where $I_\phi = \int_0^t e_\phi d\tau$ and $I_\theta = \int_0^t e_\theta d\tau$ correspond to the (unbounded) integrator states of the RL controller. \end{appendices}

\end{document}